\documentclass[twocolumn,prl]{revtex4}
\usepackage{graphicx}
\newcommand{\be}{\begin{equation}}
\newcommand{\ee}{\end{equation}}
\newcommand{\bea}{\begin{eqnarray}}
\newcommand{\eea}{\end{eqnarray}}

\newcommand{\p}{\partial}

\newcommand{\la}{\langle}
\newcommand{\ra}{\rangle}

\newcommand{\lp}{\left(}
\newcommand{\rp}{\right)}
\newcommand{\E}{{\cal E}}

\renewcommand{\phi}{\varphi}
\renewcommand{\epsilon}{\varepsilon}
\renewcommand{\vec}[1]{{\bf #1}}

\begin{document}

\title{Charge and Spin Transport at the Quantum Hall Edge of Graphene}
\author{Dmitry A. Abanin, Patrick A. Lee, Leonid S. Levitov}
\affiliation{
 Department of Physics,
 Massachusetts Institute of Technology, 77 Massachusetts Ave,
 Cambridge, MA 02139}

\begin{abstract}
Landau level bending near the edge of graphene,
described using 2d Dirac equation, provides a microscopic
framework for understanding the quantum Hall Effect (QHE) in this material.
We review properties of the QHE edge states in graphene,
with emphasis on the novel phenomena that arise due to Dirac
character of electronic states. A method of mapping out the dispersion
of edge states using scanning tunneling probes is proposed.
The Zeeman splitting of Landau levels 
is shown to create a particularly interesting 
situation around the Dirac point, 
where it gives rise to counter-circulating modes
with opposite spin. 
These chiral spin modes
lead to a rich variety of spin transport phenomena, including
spin Hall effect, spin filtering and injection, and electric detection 
of spin current. The estimated Zeeman spin gap, enhanced by exchange,
of a few hundred Kelvin, makes graphene an attractive 
system for spintronics.
Comparison to recent transport measurements near $\nu=0$
is presented.
\end{abstract}

\maketitle

\section{Introduction}

Isolation and gating \cite{Novoselov04} 
of graphene, a monolayer of graphite, has enabled the observation
of interesting transport effects, resulting from 
Dirac fermion-like character of excitations. 
In particular, graphene hosts an integer quantum Hall
effect (QHE) with unusual plateau structure\,\cite{Novoselov05,Zhang05}. 
It was found that the QHE plateaus 
in monolayer
are arranged symmetrically around the neutrality point,
occurring at filling factors which are half-integer multiples of four, 
which is the combined spin and valley degeneracy of graphite. 
This form of QHE came to be known as 
the anomalous or half-integer QHE. 

The simplest framework allowing to understand this behavior of QHE is provided
by the structure of Landau levels (LL) 
of a 2d massless Dirac equation\,\footnotemark[\thefootnote]
which has the particle-hole symmetric spectrum
%
\footnotetext{
The result (\ref{eq:bulkLL}) follows from the 
well-known representation
of the Schr\"odinger-Pauli Hamiltonian 
as a square of a massless Dirac Hamiltonian,
$H_{\rm SP}=\frac1{2m}\lp \sigma^i(p_i-\frac{e}{c}A_i)\rp^2$.
The Dirac spectrum (\ref{eq:bulkLL}) is then obtained 
by taking a square root of the nonrelativistic Landau level spectrum.
}
\be\label{eq:bulkLL}
E_n= 
\pm \epsilon_0\sqrt{|n|}
,\quad \epsilon_0=\hbar v_0 \sqrt{ 2eB/\hbar c}
\ee
with the sign plus for positive $n$, and minus for negative $n$.
Here $B$ is the magnetic field and 
$v_0\approx 8\times 10^7\, {\rm cm/s}$ 
is the velocity at the graphite Dirac point $K$ or $K'$. 
Due to spin and valley degeneracy 
(we shall discuss the role of Zeeman splitting below), 
each of the levels (\ref{eq:bulkLL}) contributes a step 
of $4e^2/h$ to the quantized Hall conductivity. 
The particle-hole symmetry $\epsilon\to -\epsilon$ of the spectrum 
(\ref{eq:bulkLL}), with the $n=0$ level positioned at $\epsilon=0$,
suggests that the QHE plateaus must occur at 
$\nu=\pm2(2n+1)=\pm 2,\pm 6,\pm 10, ...$,
which is indeed what is observed in experiment\,\cite{Novoselov05,Zhang05}.
The anomalous QHE in graphene can be understood 
in a more fundamental way in terms of a quantum anomaly 
of the zeroth Landau level\,\cite{Gusynin05}. 
The special character of the monolayer spectrum (\ref{eq:bulkLL}) 
is underscored by
the difference between the QHE properties observed in graphene 
monolayer and bilayer systems\,\cite{Novoselov06}. 

One of the most dramatic consequences of 
the Dirac LL spectrum (\ref{eq:bulkLL}) is the appearance of
the new energy scale $\epsilon_0$.
The square root dependence $\epsilon_0\propto\sqrt{B}$
leads to a much larger level spacing than that 
for electrons with the quadratic dispersion conventional for semiconductors.
For typical magnetic field of $10\,{\rm T}$, 
the separation of the lowest LL ($n=0,1$)
is quite large, $\Delta E=\epsilon_0\approx 1000 \, {\rm K}$,
which enables QHE to persist up to room temperature\,\cite{GeimKimAPS2007}.

To gain insight into the microscopic origin of the anomalous QHE,
it is useful to develop the edge-states approach,
which  provides an intuitive and simple picture of the conventional 
QHE\,\cite{Halperin}. 
The edge states for graphene were studied using
a numerical solution of the tight-binding model\,\cite{Peres} 
and also with the help of the Dirac equation\,\cite{Abanin06a,Brey06}. 
It was found that the energy levels (\ref{eq:bulkLL}), 
valley-degenerate in the graphene bulk,
are split near the edge due to valley mixing
at the boundary.
Interestingly, it turns out that the structure and dispersion 
of the edge states depend on particular
crystallographic orientation of the edge.
For the so-called armchair edge, a simple 
particle-hole symmetric splitting was found\,\cite{Abanin06a}. 
Particle-like and hole-like states have different sign of energy dispersion,
giving rise to counter-propagating modes with opposite chirality.
A somewhat more complicated situation occurs near
the zigzag edge, where dispersing edge states coexist
with an additional dispersionless surface state\,\cite{Peres}.
Despite these differences, 
the armchair, zigzag and other edges have same numbers 
of dispersing edge states
of both chiralities. As we discuss below in Sec.II\ref{sec2}, this ensures 
the universal half-integer character of QHE in graphene. 

One unique aspect of QHE in graphene is that its electronic states, 
owing to the monolayer character of this material, 
are fully exposed and, similar to the surface states of 3d materials, 
can be investigated by scanning tunneling microscopy (STM) probes with atomic resolution.
Moreover, some useful information 
can even be obtained 
by imaging the top layer of 3d graphite, as demonstrated by recent STM studies
of Dirac Landau levels\,\cite{Matsui05,Niimi06b} and of electron states 
near atomically thin edges\,\cite{Kobayashi06,Niimi06a}. 
In Sec.\ref{sec2}II we discuss new possibilities for scanning experiments that
arise in graphene. 
The characteristic spatial scale of the states in a Landau level, given
by the magnetic length $\ell_B=\left(\hbar c/eB\right)^{1/2}$,
is about $8\,{\rm nm}$ for the field of $10\,{\rm T}$.
Being large compared to the STM spatial resolution, it allows to image 
individual electronic states with sub-$\ell_B$ resolution
and, in particular, to study Landau level bending near graphene edge. 
This bending in fact mimics
the edge states momentum dispersion,
due to the position-momentum duality of the Landau levels.
As we shall see, the STM technique has sufficient resolution to map out 
the dispersion of QHE edge states.

Another novel feature of graphene is the simultaneous 
presence of the QHE edge modes of opposite chiralities, propagating in the opposite
directions. Being particle-like and hole-like, they occur at the energies
$\epsilon>0$ and $\epsilon<0$, respectively. For electron density
detuned from the neutrality point, $\nu=0$, only one of the
chiralities contributes to transport. However, as we shall see, 
near $\nu=0$ the states of both chiralities can participate in transport,
leading to rather unusual transport properties. In particular, 
in the presence of Zeeman spin splitting of Landau levels, the $\nu=0$
state features an energy gap in the bulk and, simultaneously, a pair of edge 
states of opposite chirality and opposite spin polarization\,\cite{Abanin06a,Fertig06}.
These states carry spin-up and spin-down electrons in the opposite directions
along graphene edge, exhibiting quantized spin Hall effect 
but no charge Hall effect, owing to the particle-hole symmetry at $\nu=0$.
As we discuss in Sec.III\ref{sec3}, 
these spin-polarized chiral edge states exhibit interesting 
spin transport phenomena,
such as spin filtering and spin injection, whereas the spin Hall effect
provides a natural tool for the detection of spin current.

Interestingly, 
the counter-propagating edge states
manifest themselves directly in charge transport.
As we discuss in Sec.IV\ref{sec4},
near $\nu=0$ longitudinal resistivity remains finite, 
$\rho_{xx}
\gtrsim h/e^2$, while the Hall effect, 
which is absent at the particle-hole symmetry point $\nu=0$,
appears at nonzero $\nu$ due to conductivity in the bulk. This leads to
Hall resistance $\rho_{xy}$ 
changing sign at $\nu=0$ without exhibiting a clear plateau.
The bulk conductivity short-circuits the edge transport and suppresses
longitudinal resistivity, leading to a prominent peak in $\rho_{xx}$ near 
$\nu=0$ and a plateau in $\sigma_{xy}$ 
These predictions, as well as the behavior of resistance fluctuations,
which are enhanced near zero $\nu$, 
are in agreement with experiment\,\cite{Zhang06,Abanin07}.

Looking beyond graphene, interesting massless Dirac fermion states 
have been predicted a while ago\,\cite{Volkov85} 
at interfaces of narrow-gapped HgTe and PbTe semiconductors.
In Ref.\cite{Volkov85}, which anticipated many of the features 
of electronic states in graphene, 
2d Dirac states occur in a band-inverted heterojunction plane
due to spin orbit
interaction in 3d bulk, in the absence of magnetic field.

We note also that there have been interesting predictions 
of quantized spin Hall effect in certain 2d insulators\,\cite{KaneMele,Bernevig06}. In particular, the proposal of Ref.\cite{Bernevig06} can be viewed
as a 2d version of the 3d situation discussed in Ref.\cite{Volkov85}.
In these schemes an energy gap forms in the bulk due to spin-orbit 
even in the absence of magnetic field, similar to Ref.\cite{Volkov85}.
At the same time, counter-propagating
modes carrying opposite spins appear on the edge, which are responsible 
for the quantized spin Hall current. 
%
As was emphasized in Ref.\cite{Fu06}, there are general symmetry 
requirements, rooted in the time-reversal symmetry,
protecting counter-propagating gapless excitations 
at an insulator boundary. In particular, certain $Z_2$ 
invariants must exist, which are realized as spin $\sigma_z$ component
in the case of graphene\,\cite{KaneMele,Abanin06a}, 
and were linked to supersymmetry 
for heterojunction systems\,\cite{Volkov85}.
 


We note that the nature of the edge states 
of Refs.\cite{KaneMele,Bernevig06}
is essentially identical to ours. 
The only difference, apart from different 
size of the energy gap, is that 
Rashba spin-orbit may
cause backscattering in our case (see Sec.III)
but not in the situation  of Refs.\cite{Volkov85,KaneMele,Bernevig06}, while
magnetic impurities will cause backscattering in both cases.
In particular, 
our discussion in Sec.IV of $\rho_{xx}$ and $\rho_{xy}$ and our proposals to operate and 
detect spin current can serve as diagnostic tools should these schemes become 
realized experimentally. 

The rest of the paper is organized as follows. 
In Sec.II we introduce the edge states 
using the Dirac equation framework.
We focus on the two main edge types,
armchair and zigzag, however we emphasize the generic aspects 
that are applicable to other edges. 
Then we discuss the anomalous QH effect 
and the possibility of imaging the edge states
with scanning tunneling probes. 
In Sec.III we focus on spin-polarized chiral edge states 
and related spin transport phenomena. 
We also comment on spin relaxation mechanisms
and present estimates of the spin relaxation time.
In Sec.IV we introduce a transport model which accounts for 
both edge and bulk transport. This allows us to connect 
the chiral spin-polarized edge picture with
recent transport measurements near the neutrality point\,\cite{Zhang06,Abanin07}.

\section{II. Dirac QHE edge states}
\label{sec2}

\begin{figure}
\includegraphics[width=3.1in]{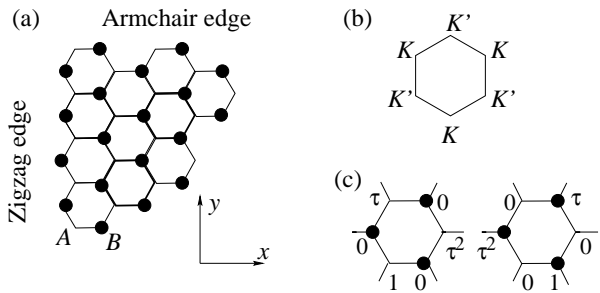}
\caption[]{(a) Graphene lattice with armchair and zigzag edges. 
Sublattices $A$ and $B$ are marked.
(b) Graphene 
hexagonal Brillouin zone with Dirac valleys $K$ and $K'$.
(c) The two linearly independent zero-energy 
Bloch functions for the $K$ valley used in (\ref{eq:dirac_hamiltonian}), 
where $\tau= e^{2\pi i/3}$. 
For the $K'$ point,
the zero-energy Bloch functions
are obtained from those shown by complex conjugation.}
\label{fig0}
\end{figure}

Here we analyze electron states 
near zigzag and armchair edges (Fig.\ref{fig0}a), the two most common 
graphene edge types, using massless Dirac model\,\cite{diVincenzo}. 
This exercise, which amounts
to setting the boundary conditions for the Dirac spinor 
and solving an appropriate 1d eigenvalue problem, provides a fully microscopic
picture of graphene QHE. 
Below we use this approach to illustrate the interplay between 
the QHE edge states and surface states 
for the zigzag edge, and discuss the possibility of imaging the edge
states with scanning tunneling probes. 


Let us first recall how low-energy graphene excitations
are obtained in the tight-binding model\,\cite{diVincenzo} 
near the Dirac valleys $K$ and $K'$, located at the two non-equivalent 
Brillouin zone corners (see Fig.\ref{fig0}(b)). 
There are two linearly independent zero-energy Bloch functions 
for each of the points $K$, $K'$, each residing only on one sublattice 
($A$ or $B$)
and vanishing on the other sublattice.
Our choice of Bloch functions for the $K$ valley 
is shown in Fig.\ref{fig0}(c). The Bloch functions 
for the valleys $K$ and $K'$ are related by complex conjugation. 


The wave function of
low-lying excitations near $K$ and $ K'$, 
is written as a superposition of these four zero-energy Bloch functions
multiplied by slowly 
varying envelope functions $u_K$, $v_K$, 
$-u_{K'}$, $-v_{K'}$,
with $u$ and $v$ being the wave function amplitudes on the sublattice $A$ 
and $B$. 
(Our choice of the signs for the $K'$ valley is 
convenient for treating an armchair boundary, 
as we shall see below.) 
The envelope functions $u_K$, $v_K$, and $u_{K'}$, $v_{K'}$ 
describe excitations near $K$ and $K'$, respectively.
The effective low-energy Hamiltonian, obtained by keeping only 
lowest-order gradients of $u$ and $v$, takes the massless Dirac form
\footnotemark:
\footnotetext{We note that the Hamiltonian used in our
paper \cite{Abanin06a} is related to the Hamiltonian 
(\ref{eq:dirac_hamiltonian}) 
by variable transformation $u_{K,K'}\to u_{K,K'}$, 
$v_{K,K'}\to -iv_{K,K'}$.}
\be\label{eq:dirac_hamiltonian}
H_K=i v_0 
 \left[\begin{array}{cc}
         0 & \tilde{p}_+\\
         -\tilde{p}_-& 0
      \end{array}
 \right], \,\,\,\,
H_{K'}=iv_0
  \left[\begin{array}{cc}
         0 & \tilde{p}_- \\
         -\tilde{p}_+ & 0 
 \end{array}\right], 
\ee
where $\tilde{p}_{\pm}=\tilde{p}_x\pm i\tilde{p}_y$,  
$\tilde{p}_{\mu}=p_{\mu}-(e/c)A_{\mu}$.

\begin{figure}
\includegraphics[width=1.64in]{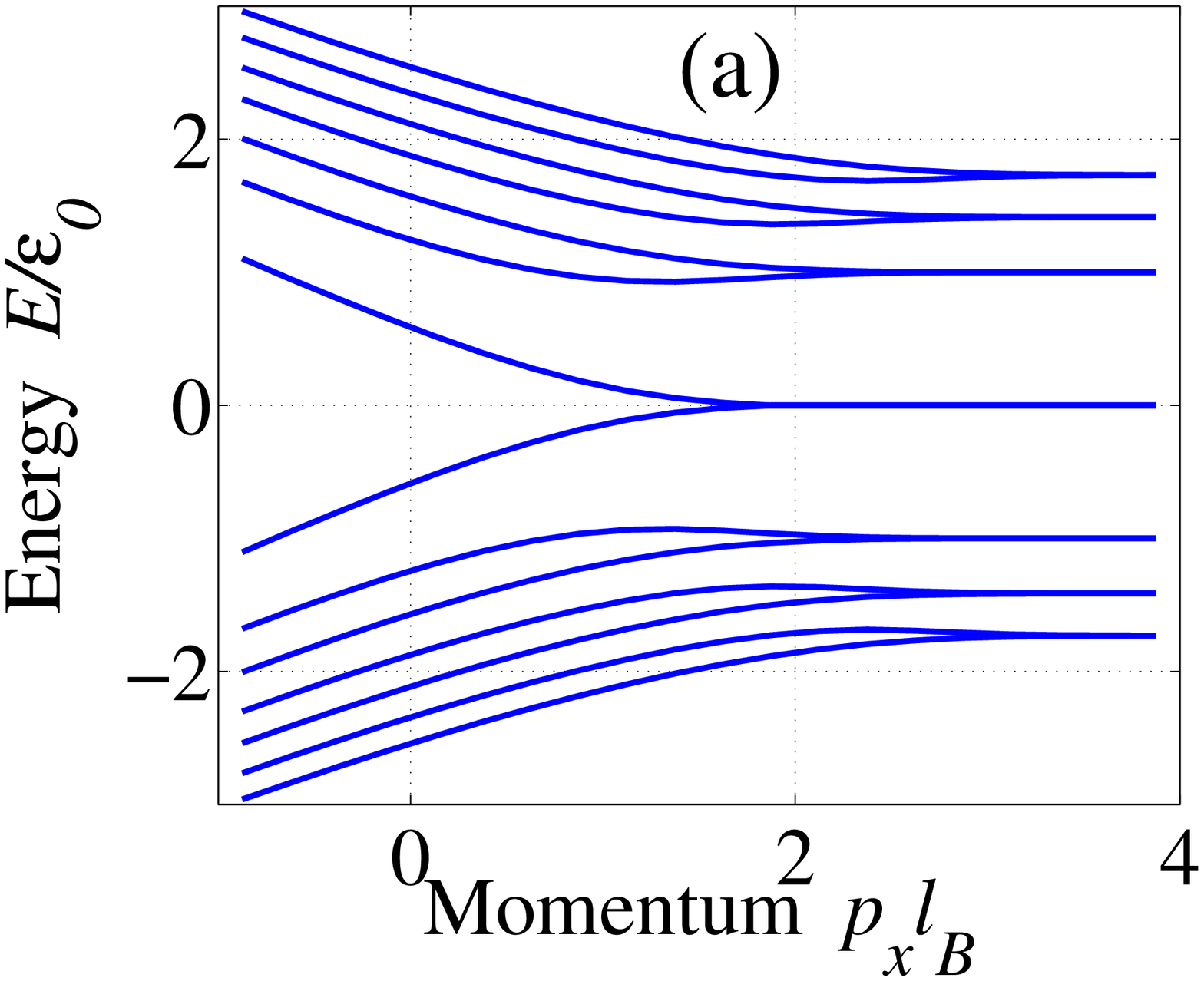}
\includegraphics[width=1.64in]{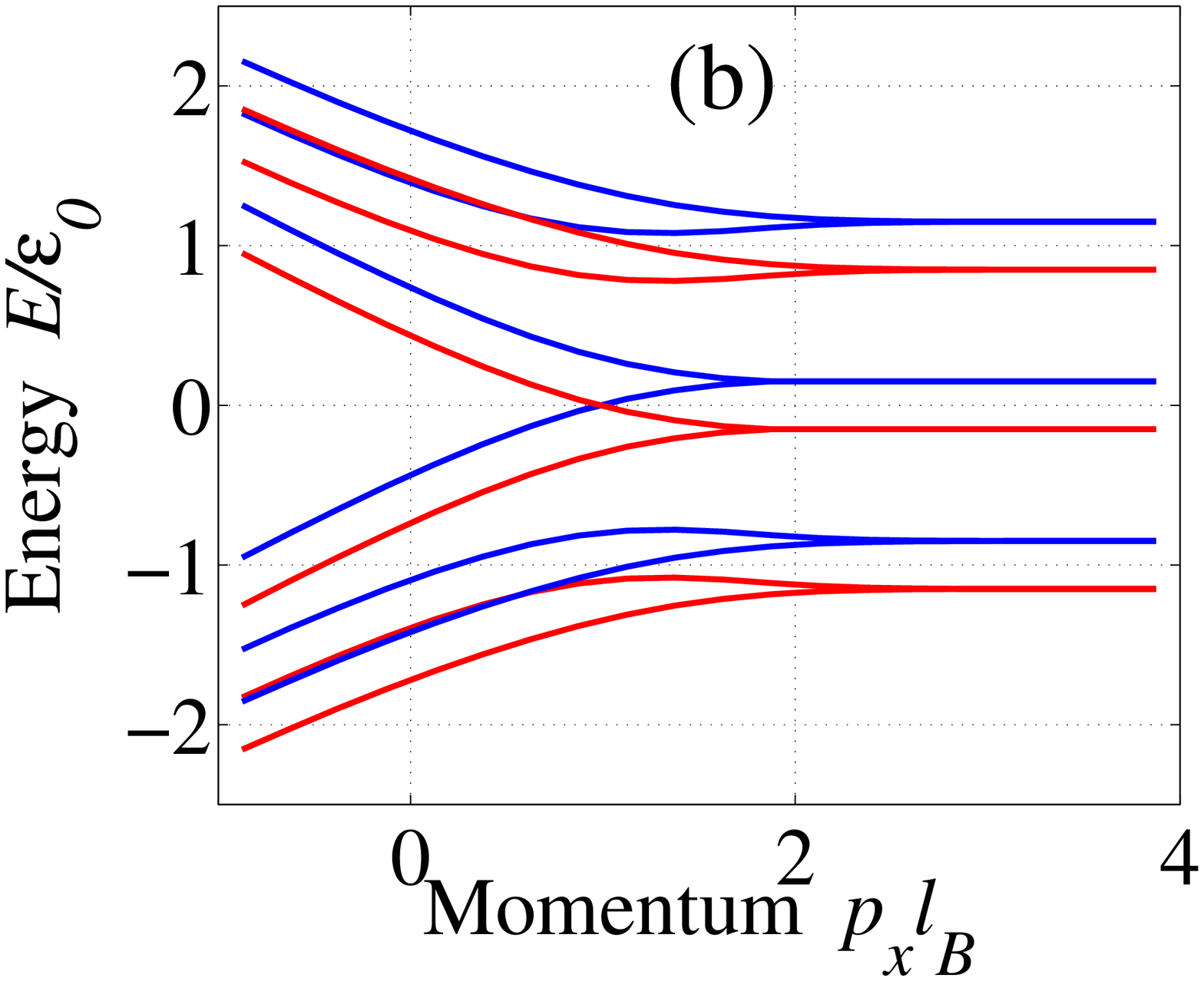}
\vspace{0.15cm}
\caption[]{(a) Electron energy spectrum near the armchair boundary
obtained from the Dirac model, Eq.(\ref{eq:dirac_hamiltonian}). 
The boundary condition, Eq.(\ref{eq:armchair_boundary}),
lifts the $K$, $K'$ degeneracy, forming particle-like 
and hole-like counter-propagating edge modes. 
The odd integer numbers of the modes yields
the ``half-integer'' QHE. 
(b) Spin-split graphene edge states, propagating in opposite directions at zero energy: 
the blue (red) curves represent the spin up (spin down) states. 
}
\label{fig1}
\end{figure}

Landau levels in an
external $B$-field, described with the  gauge
$ A_x=-By,\,  A_y=0$, 
can be obtained for the states with the $x$ dependence $e^{ip_xx}$
from 1d Hamiltonians 
%
\be\label{eq:hamiltonian_magnetic}
H_{K,K'}=\frac{i\epsilon_0}{\sqrt{2}} 
 \left[\begin{array}{cc}
         0 & \pm \partial_y+(y-y_\ast)\\
         \pm \partial_y-(y-y_\ast)& 0
      \end{array}
 \right], 
\ee
where $\epsilon_0=\hbar v_0\lp 2eB/\hbar c\rp^{1/2}$ and $y_\ast=-p_x$. 
Here $y$ and $p_x$ are measured in the units
of $\ell_B$ and $\hbar/\ell_B$, respectively. The spectrum of $H_{K,K'}$, Eq.(\ref{eq:hamiltonian_magnetic}),
yields the Dirac Landau levels,
Eq.(\ref{eq:bulkLL}),
where the eigenstates
for the two valleys $K(K')$ are given by 
%
%
%
\bea\label{eq:zeroth_LL_K}
&& \lp u_{K,n},v_{K,n}\rp =A\lp c_n\phi_{n-1}(y-y_\ast),\phi_n(y-y_\ast)\rp,
\\ \label{eq:zeroth_LL_K'}
&& \lp u_{K',n},v_{K',n}\rp =A\lp \phi_n(y-y_\ast),c_n\phi_{n-1}(y-y_\ast)\rp
.\
\eea
%
Here $\phi_n(z)$ 
are the eigenfunctions of the magnetic oscillator, $n=0,1,...$, 
the normalization factor $A$ equals 1 for $n=0$ and $1/\sqrt{2}$ for 
$n\neq 0$, with $c_0=0$ and $c_{n\ne0}=1$.
Note that zeroth LL states reside solely on $B$ sublattice
for valley $K$ and on $A$ sublattice for valley $K'$
\footnotemark.
\footnotetext{ 
This property is specific for the zeroth LL,  
making the splitting of the $n=0$ LL due to Coulomb interaction 
distinctly different from that of other LLs (see \cite{Abanin06b,Goerbig06}).}

We now analyze how LL spectrum is modified near the {\it armchair} edge.  
We consider graphene sheet in the halfplane $y<0$ 
with an armchair edge parallel to the 
$x$ axis (see Fig.\ref{fig0}(a)). 
Energy levels 
near the edge are determined from the Dirac
eigenvalue equations 
$E\psi=H_{K,K'}\psi$, where $\psi=(u,v)$, and $H_{K,K'}$ are
given by Eq.({\ref{eq:hamiltonian_magnetic}}).
To analyze this eigenvalue problem,
we exclude $v$ components 
and consider eigenvalue equations 
with spectral parameter $\lambda=(E/\epsilon_0)^2$
for $u$ components:
\be\label{eq:equations_uK_comp}
\begin{array}{r}
\frac12\left(-\partial _y^2+(y-y_\ast)^2+1  \right)u_K=\lambda u_K, 
\\
\frac12\left(-\partial_y^2+
(y-y_\ast)^2-1\right)u_{K'}=\lambda u_{K'}, 
\end{array}
\ee
%
%
The boundary conditions for 
Eqs.(\ref{eq:equations_uK_comp}) can be obtained 
from the tight-binding model, which 
is valid up to the very last row near $y=0$,
by setting the 
wave function equal zero at the boundary.
Since the armchair edge has lattice sites of both $A$ and $B$ type (see
Fig.\ref{fig0}(a)), 
the wave function  
on both sublattices should vanish at the edge. In terms of the
envelope functions $u_{K,K'}$, $v_{K,K'}$, taken at $y=0$, this condition 
translates into 
\be\label{eq:armchair_boundary}
u_K=u_{K'},\quad v_K=v_{K'}. 
\ee
%
%
We obtain a pair of differential equations (\ref{eq:equations_uK_comp})
on the semi-axis $y<0$, coupled at the boundary via Eq.(\ref{eq:armchair_boundary}). 
To simplify this problem, let us map Eq.(\ref{eq:equations_uK_comp}) 
for $u_{K'}$
onto the positive 
semi-axis $y>0$, by $u_{K'}(y)\to u_{K'}(-y)$, 
while keeping $u_K$
on the negative semi-axis $y<0$,
and treat it as an eigenvalue problem in the domain $-\infty<y<\infty$
with the wavefunction given by $u_K$ at negative $y$ and by $u_{K'}$
at positive $y$.  
The first boundary condition, Eq.(\ref{eq:armchair_boundary}), 
then means that the wavefunction is continuous at $y=0$, 
while the second condition implies continuity of the derivative 
$\partial u/\partial y$. (This can be seen by expressing $v_{K,K'}$
in terms of $u_{K,K'}$ using Eqs.(\ref{eq:hamiltonian_magnetic}).)
Thus we obtain a 1d Schr\"odinger problem 
in the potential\,\cite{Abanin06a}
\be\label{eq:armchair_potential}
V(y)=\frac12(|y|+y_\ast)^2-\frac12{\rm sgn}(y),
\ee
defined on the entire $y$ axis. 
After finding the spectrum $\lambda(y_\ast)$ numerically, we obtain
the energy levels of the Dirac fermions as
%
\be\label{eq:E_lambda} 
E(p_x)=\pm \epsilon_0 \sqrt{\lambda(y_\ast)}
,\quad
y_\ast=-p_x
, 
\ee
whereby the particle-hole symmetry is restored 
due to the two possible signs of $E(p_x)$. 
The dispersion (\ref{eq:E_lambda}) is 
illustrated in Fig.\ref{fig1}a. 
We note that the LL double valley 
degeneracy in the bulk is lifted at the boundary. 

The particle-hole symmetric  edge states spectrum in Fig.\ref{fig1}a
instantly explains the ``half-integer" Hall quantization in graphene. 
Indeed, for any electron density with integer $\nu$ in the bulk 
there is an odd 
number of 
the edge modes crossing the Fermi level, which means that
the Hall conductivity is quantized as $\sigma_{xy}=2(2n+1)e^2/h$,
where the factor two accounts for spin degeracy.


Spin degeracy of the Landau levels is lifted 
by the Zeeman interaction, which is substantial in 
graphene, 
%
%
\be\label{eq:zeeman}
E_Z= g\mu_B B\approx 50\, {\rm K}
,\quad g\approx 2
,
\ee
%
for $B = 30\,{\rm T}$ (compare to $g\sim 0.1$ in GaAs quantum wells). 
Zeeman-split edge states, depicted in Fig.\ref{fig1}b, have interesting
characteristics
for electron density near neutrality, 
$\nu=0$. At this density the state in the bulk is spin-polarized,
with the Zeeman gap further enhanced by exchange (see below).
The two branches of counterpropagating 
edge states near $\epsilon=0$, carrying opposite spin, 
have interesting properties that will be 
discussed in more detail in Sec.III. 

We now analyze the {\it zigzag} edge, which  
even at $B=0$ hosts a band of dispersionless 
zero-energy states 
bound to the edge\,\cite{Fujita}. We shall refer to these states
as surface states, 
to distinguish them from the dispersing QHE edge states. 
As we shall see, the surface states contribute 
to the splitting of $n=0$ LL near the zigzag edge,
in agreement with the tight-binding calculations\,\cite{Peres}.
 
\begin{figure}
\includegraphics[width=1.64in]{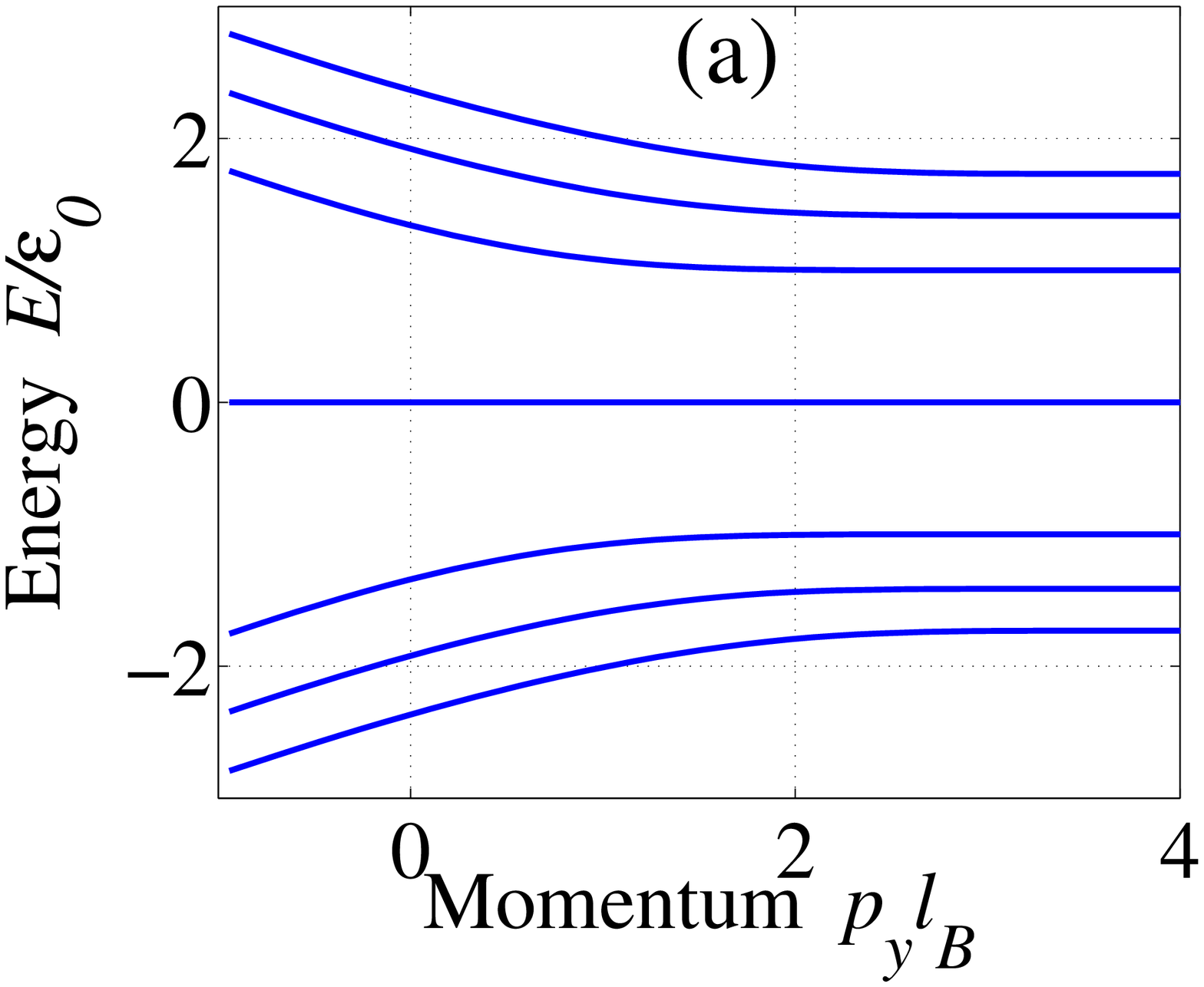}
\includegraphics[width=1.64in]{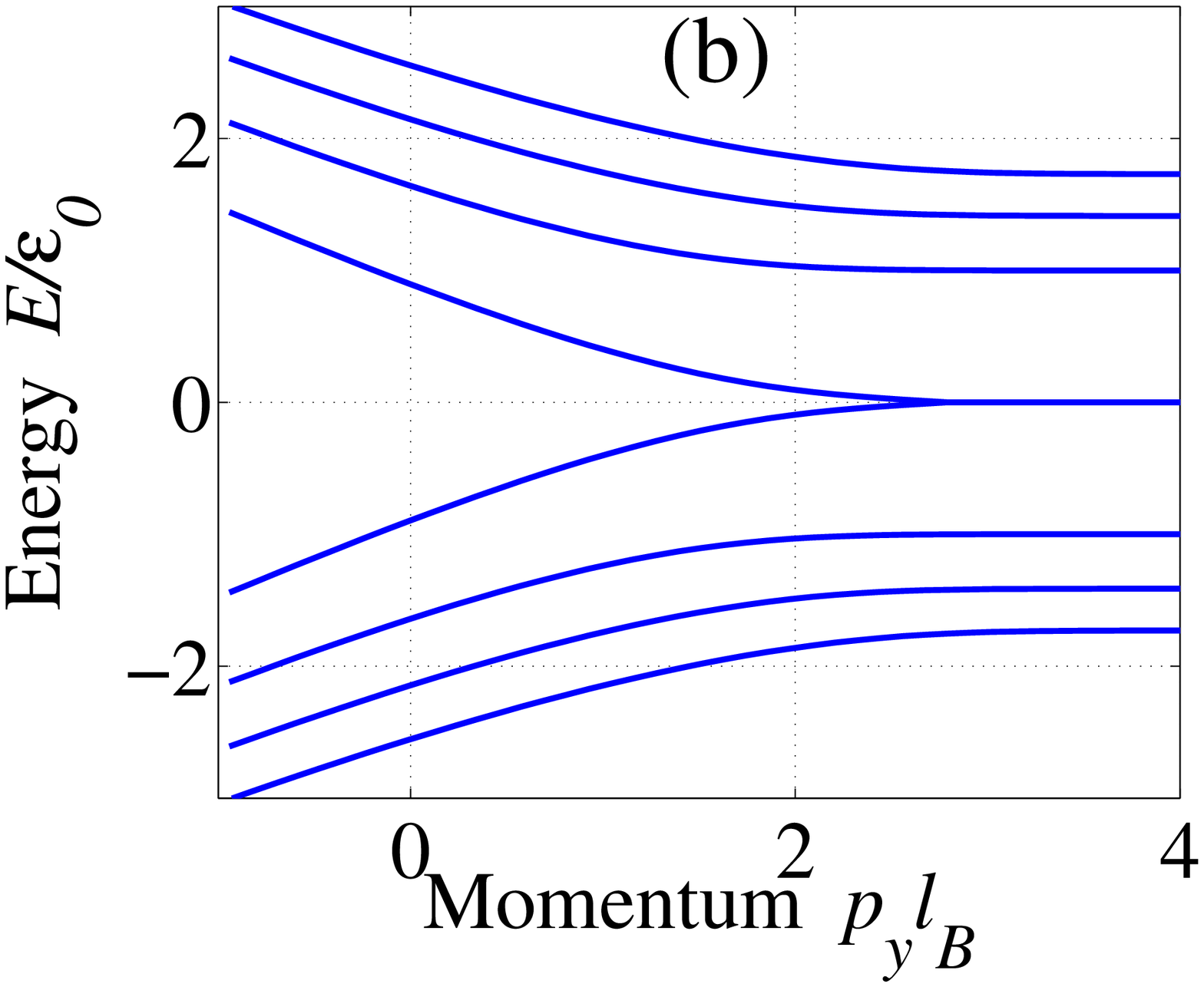}
\vspace{0.15cm}
\caption[]{Electron energy dispersion for the zigzag boundary 
obtained from the Dirac model, Eq.(\ref{eq:dirac_hamiltonian}), (\ref{eq:boundary_zigzag}). (a) Spectrum for 
the $K$ valley. The zeroth LL morphs into dispersionless 
surface mode\,\cite{Fujita} near the edge.
(b) Spectrum for the $K'$ valley. The zeroth LL mixes with 
the surface mode, giving rise to two branches of
dispersing QHE edge states.}
\label{fig3}
\end{figure}

We consider graphene sheet
in the half-plane $x>0$, with its first row consisting of $B$ 
atoms (see Fig.\ref{fig0}a). 
For the states with the $y$ dependence $e^{ip_yy}$,
with $ A_x=0$, $A_y=Bx$, 
from (\ref{eq:dirac_hamiltonian}) we obtain
1d Hamiltonians
%
\be\label{eq:hamiltonian_magnetic_zigzag}
H_{K,K'}=\frac{\epsilon_0}{\sqrt{2}} 
 \left[\begin{array}{cc}
         0 &  \partial_x\pm(x-x_\ast)\\
         -\partial_x\pm (x-x_\ast)& 0
      \end{array}
 \right], 
\ee
where $x_\ast=p_y$. 
Similarly to the armchair case, the spectrum can be found from the eigenvalue 
equation $E\psi=H_{K,K'}\psi$, where $\psi=(u,v)$, which should be supplemented
with the boundary conditions. For our zigzag edge 
the wavefunction have must vanish on all $A$ sites at $x=0$.
%
%
For that both envelope functions $u_K, u_{K'}$ have to vanish 
at the boundary, 
\be\label{eq:boundary_zigzag}
u_K=0,\quad u_{K'}=0.
\ee
Excluding $v$ components,
we obtain two separate 
eigenvalue problems for the spectral parameter $\lambda=(E/\epsilon_0)^2$,
\be\label{eq:uK_comp_zigzag}
\begin{array}{r}
\frac12\left(-\partial_x^2+(x-x_\ast)^2+1  \right)u_K=\lambda u_K, 
\\
\frac12\left(-\partial_x^2+(x-x_\ast)^2-1\right)u_{K'}=\lambda u_{K'},
\end{array}
\ee
where both $u_K$ and $u_{K'}$ 
satisfy the hard wall boundary conditions 
(\ref{eq:boundary_zigzag}).
The amplitudes $v_{K,K'}$ on the $B$ sublattice can be 
expressed via ampliudes $u_{K,K'}$ on the $A$ sublattice and eigenenergy $E$,
\be
\label{eq:vK_uK}
\begin{array}{r}
v_{K}=(\epsilon_0/\sqrt{2}E)\left(-\partial_x+(x-x_\ast)\right)u_K,
\\
v_{K'}=(\epsilon_0/\sqrt{2}E)\left(-\partial_x-(x-x_\ast)\right)u_{K'}.
\end{array}
\ee
The eigenvalue problems (\ref{eq:uK_comp_zigzag}) 
with the hard-wall boundary conditions (\ref{eq:boundary_zigzag}) are familiar 
from the theory of edge states in the conventional QHE\,\cite{Halperin}, and their 
spectrum $\lambda(x_\ast)$ can be found numerically. 
The Dirac fermion energy dispersion $E(p_y)=\pm\epsilon_0\sqrt{\lambda(x_\ast)}$  
is shown in Fig.\ref{fig3}.

The behavior of $n\neq 0$ LL's is similar to the armchair case:
there are two 
branches of the edge states, one for each valley, 
degenerate in the bulk, $x_\ast\gg1$, which split near the edge. 
The zeroth LL, however, coexists with the surface state,
which makes its behavior rather peculiar and different for the two valleys.

In the $K$ valley, which we discuss first,
the zeroth LL $K$ states reside solely on the 
$B$ sublattice, see Eq.(\ref{eq:zeroth_LL_K}), 
and therefore automatically satisfy the boundary condition 
$u_K=0$. Thus there are zero-energy states for arbitrary 
values of $x_\ast$, of the form $\phi_0(x-x_\ast)\propto e^{-(x-x_\ast)^2/2}$.
Let us consider the states with $x_\ast$ far outside the graphene
half-plane, $x_\ast\ll -1$. Not too far from the boundary,
such states can be approximated by an exponential
%
\be\label{eq:K_zero_energy}
v_{K}(0<x\lesssim |x_\ast|)\propto e^{-|x_\ast|x}
,\quad u_{K}(x)=0.
\ee
which is identical to surface state wave function \cite{Fujita}. 
Thus the zeroth LL for valley $K$ near the edge
transforms into the surface mode. 
Being dispersionless, this mode
does not contribute to the edge transport.
The edge state spectrum for the valley $K$ is displayed in Fig.\ref{fig3}a. 

Now let us consider the zeroth LL for the valley $K'$.
For $x_\ast\gg 1$, we approximate the ground state of the 
oscillator (\ref{eq:uK_comp_zigzag}) with the 
hard-wall boundary condition as
\be\label{eq:oscillator_antisymm}
u_{K'}(x)=\psi_{x_\ast}(x)\approx \phi_0(x-x_\ast)-\phi_0(x+x_\ast)
.
\ee
%
The ground state energy $\lambda_0(x_\ast)$ 
is then approximated by 
\be\label{eq:GS_energy}
\lambda_0(x_\ast)\approx \la h \ra,
\ee 
where $h=\frac{1}{2}(-\partial_x^2+(x-x_\ast)^2-1)$ is the effective 
Hamiltonian for $u_{K'}$ 
component, Eq.(\ref{eq:uK_comp_zigzag}), and $\la ...\ra$ 
denotes averaging over the 
normalized wave function (\ref{eq:oscillator_antisymm}). 

Evaluating $\la h \ra$ for $x_\ast\gg 1$,
when the state 
(\ref{eq:oscillator_antisymm}) has unit norm with exponential accuracy,
we obtain
\be\label{eq:lambda_approx}
\lambda_0(x_\ast)\approx x_\ast \pi^{-1/2} e^{-x_\ast^2}.
\ee
From the relation $E=\pm\epsilon_0\sqrt{\lambda}$,
we find the energies for the two branches 
of dispersing edge states
%
\be\label{eq:GS_energy_appr}
E_\pm (x_\ast)\approx \pm (2x_\ast)^{1/2}\pi^{-1/4} e^{-{x_\ast^2}/2} \epsilon_0. 
\ee
Plugging this expression in Eq.(\ref{eq:vK_uK}), 
we obtain the wave function on $B$ sublattice for these two branches, 
%
\be\label{eq:vK'_GS}
v_{K'}=\pm x_\ast^{1/2} e^{-x_\ast x}. 
\ee
%
which is again the surface state wave function 
(compare to Eq.(\ref{eq:K_zero_energy})).   
We therefore conclude that for the $K'$ valley 
the zeroth Landau level and the surface state mix 
giving rise to two dispersing edge modes. 
This is in agreement with the spectrum displayed in Fig.\ref{fig3}b. 
We see that, although the Dirac model is applicable
only in a small part of the Brilloin zone, near points $K$ and $K'$,
it provides a description of the states at the zigzag edge, including the
surface state, which is in agreement with the results of the tight-binding 
model of Ref.\cite{Peres}. 
The surface mode in the vicinity of 
$K$ and $K'$ is given by Eqs.(\ref{eq:K_zero_energy}),(\ref{eq:vK'_GS}).




Interestingly, 
the $A$ and $B$ sites contribute equally
to the splitting of the zeroth LL, 
$\int_{x>0} |u_{K'}|^2dx = \int_{x>0} |v_{K'}|^2dx$.
This is somewhat counterintuitive, since this LL
is solely on the sublattice $A$ in the bulk, while the surface mode is solely on
the sublattice $B$. 
This equal participation property can be understood as follows.
The spinor states $(u_{K'},\pm v_{K'})$ with $u_{K'}$, $v_{K'}$
given by Eqs.(\ref{eq:oscillator_antisymm}),(\ref{eq:vK'_GS}), are eigenstates
of the Dirac Hamiltonian with the boundary condition 
(\ref{eq:boundary_zigzag}), 
with the energies
$\pm E$. Thus these states are orthogonal, which implies
that the integrals of $|u_{K'}|^2$ and $|v_{K'}|^2$ are equal.
We further note that the integral of the square of the $B$ component of
our edge state wave function (\ref{eq:vK'_GS}) 
over $x>0$ indeed equals one, in agreement with our choice of normalization
on the $A$ sublattice in Eq.(\ref{eq:oscillator_antisymm}).

To sum up, for the zigzag edge, the zeroth LL  gives rise to two 
dispersing edge states for one of the valleys, while for the other valley 
the zeroth LL morphs into the dispersionless surface mode 
which does not contribute to the edge current. 
Therefore, despite the presence 
of the surface mode, the number 
of dispersing QHE edge states with $\epsilon>0$ and $\epsilon<0$
for the zigzag boundary 
is the same as for the armchair boundary, giving rise to
``half-integer" quantization of Hall current.


\begin{figure}
\includegraphics[width=1.66in]{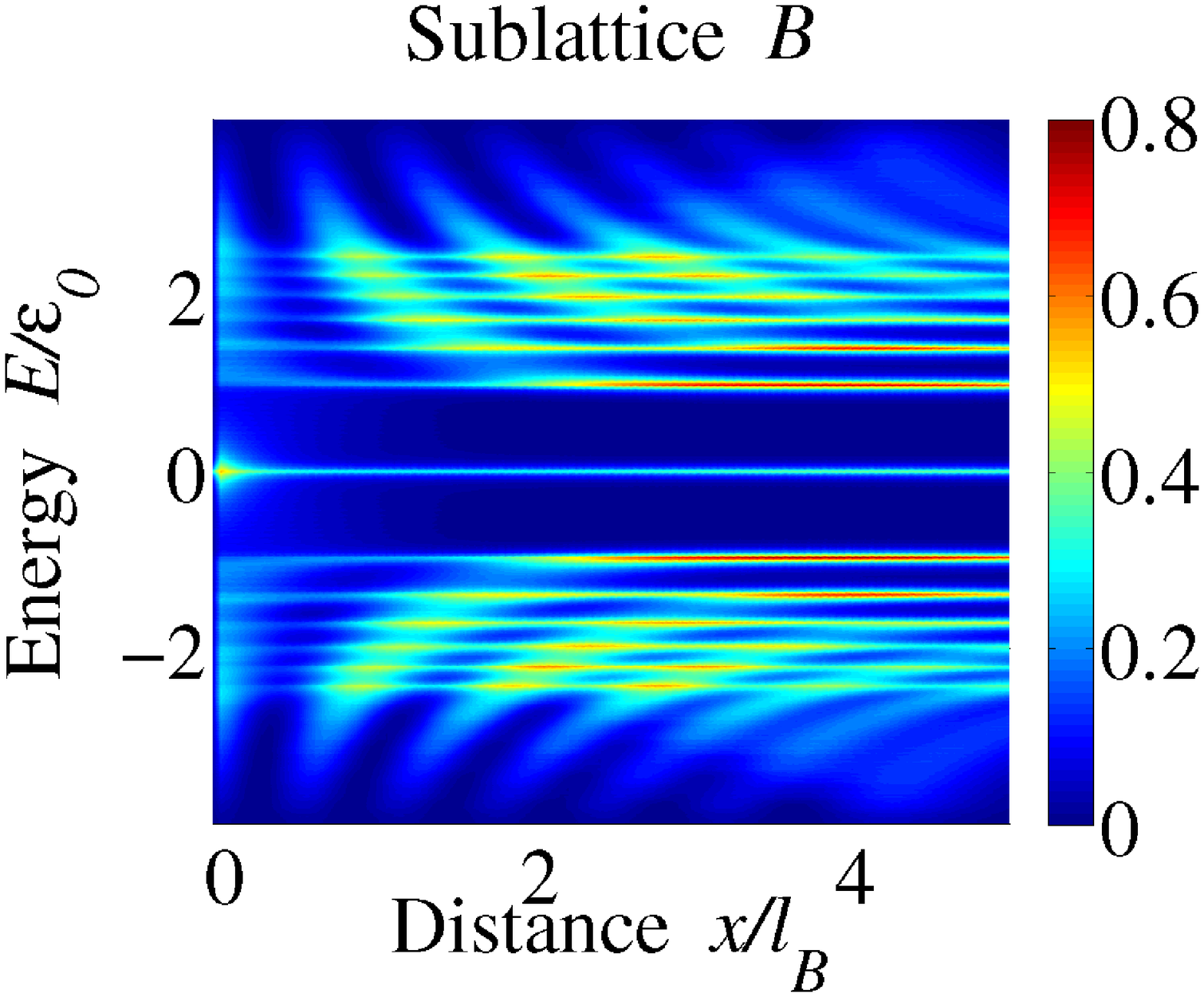}
\includegraphics[width=1.66in]{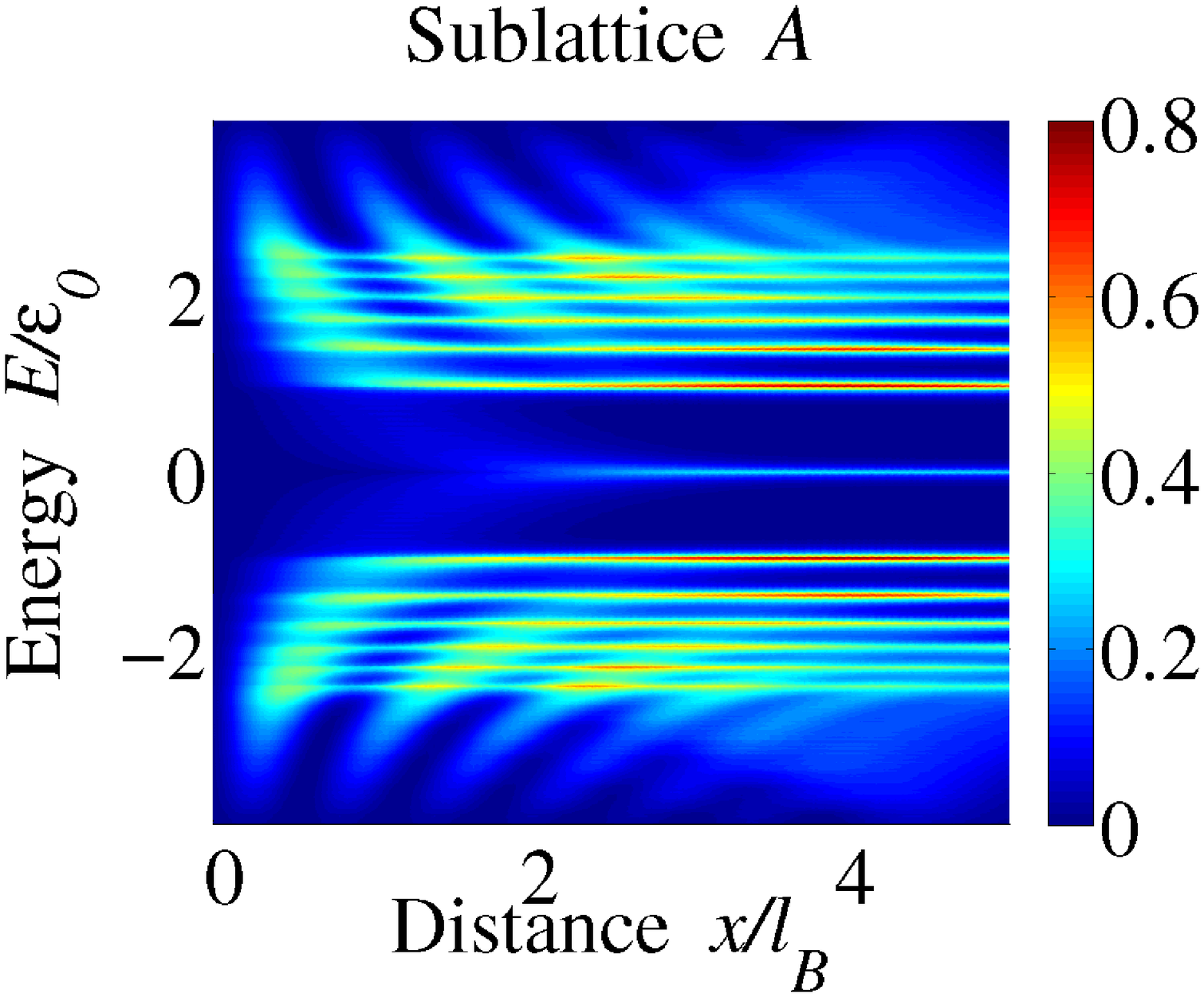}
\vspace{0.15cm}
\caption[]{Position-dependent tunneling spectroscopy of graphene 
near the zigzag edge for sublattices $A$ 
and $B$. 
Due to the momentum-position 
duality, the dependence of LDOS, Eqs.(\ref{eq:LDOS_A}), 
on the the distance to the edge $x$
mimics the momentum dispersion of the edge states.
Note the difference of the spectra
for the two sublattices near the edge and presence of 
surface state for sublattice $B$.
}
\label{fig6}
\end{figure}

Finally, we briefly discuss how the 
edge states in graphene  
can be investigated using the STM
technique\,\cite{Matsui05,Niimi06b,Kobayashi06,Niimi06a}. 
Due to the Landau level momentum-position duality relation,
$p_y=(\hbar/\ell_B^2)x_\ast$,
the edge state momentum dispersion shown in Figs.\ref{fig1},\ref{fig3}
translates into the excitation energy dependence 
on the distance from the edge.
The characteristic scale for the latter is set by the magnetic
length $\ell_B$, which for typical fields is about 50-80 times greater than 
the spatial resolution of STM instruments on graphite surface.
This makes STM technique particularly convenient for this kind of
studies.



A link between the edge states dispersion and the position-dependent 
tunneling spectroscopy can be established as follows.
We shall use the solutions for the edge state wave function 
given above to calculate 
the local density of states (LDOS) near the zigzag edge
(other edge types can be dealt with similarly). 
For each of the graphene sublattices 
LDOS is given by 
\be\label{eq:LDOS_A}
\begin{array}{l}
\rho_{A}(E,x)=\sum_{\alpha} \left| u_{\alpha}(x)\right|^2 \delta(E-E_{\alpha}),
\\
\rho_{B}(E,x)=\sum_{\alpha} \left| v_{\alpha}(x)\right|^2 \delta(E-E_{\alpha}),
\end{array}
\ee
where $x$ is the distance from the edge, 
and $\alpha$ denotes the set of eigenstates of 
the $K$ and $K'$ Hamiltonians (\ref{eq:hamiltonian_magnetic_zigzag}) 
with the hard-wall boundary condition (\ref{eq:boundary_zigzag}).



Using the eigenfunctions $u_{\alpha}(x)$, $v_{\alpha}(x)$ and the energies
$E_{\alpha}$ found from Eqs.(\ref{eq:uK_comp_zigzag}),(\ref{eq:vK_uK}) 
as discussed above, we obtain LDOS for the $A$ and $B$ sublattices 
which is displayed in Fig.\ref{fig6}.
We see that the position-independent Landau level bands, dominating LDOS 
far from the edge, bend away from $\epsilon=0$ near the edge.
This bending mimics the edge states momentum dispersion
shown in Fig.\ref{fig3}. Note, however, that LDOS is nonzero
only for $x>0$, whereas the edge state momentum $p_y$ can be both positive
and negative. The spatial width of the bending bands is determined
by the width of the eigenfunctions $u_{\alpha}(x)$, $v_{\alpha}(x)$,
which is of the magnetic length scale.

\section{III. Spin-polarized chiral edge states and spin transport.}
\label{sec3}


As we noted above, at the neutrality point $\nu=0$ graphene hosts gapless
spin-polarized edge states (see Fig.\ref{fig1}(b)).
The Zeeman energy gap in the bulk,
Eq.(\ref{eq:zeeman}), 
is enhanced by the Coulomb interaction. 
A Hartree-Fock estimate of this enhancement\,\cite{Abanin06a} gives 
%
\be\label{eq:spin_gap_ex}
\Delta =\frac{\pi^{1/2}e^2}{2\kappa\hbar v_0} (1-\alpha)\epsilon_0
\approx 0.456\cdot (1-\alpha)  \epsilon_0,
\ee
where 
$\kappa = 1+ \pi e^2/{2\hbar v_0} \approx 5.24$ is RPA 
screening function, and the parameter $0<\alpha<1$ describes 
relative strength of Coulomb and exchange correlations. Assuming $\alpha=0$,
i.e. ignoring correlations of electrons with opposite spin, 
we obtain a spin gap $\Delta\sim 800\, {\rm K}$ for $B= 30\,{\rm T}$. 
Taking into account the substrate dielectric constant, $e^2\to\frac2{\epsilon+1}e^2$, changes the result only slightly 
($\frac2{\epsilon+1}=0.36$ for ${\rm SiO_2}$).
This approximation, 
while pointing at a correct order of magnitude of a few hundred Kelvin,
probably somewhat overestimates the spin gap since it ignores
correlations and disorder effects.


The chiral spin-polarized 
edge states offer a unique setting to study spin transport.
In particular, the spin-split state $\nu=0$ may be 
used to generate and detect spin-polarized currents. 
This spin transport regime
seems attractive due to the large bulk gap and 
high stability of the edge states. Moreover, 
increased quality of samples should allow 
existence of spin polarized edge states even at relatively low magnetic fields.

\begin{figure}
\includegraphics[width=3.1in]{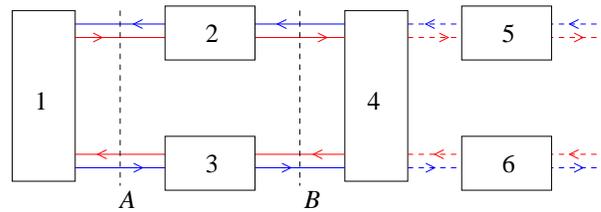}
\caption[]{
A Hall bar at $\nu=0$ can be used to generate and detect spin currents.
Blue and red lines represent edge currents with up and down spins. 
Contacts 1 and 4 
are source and drain, which may be used to inject spin polarized current.
Contacts 2, 3 are voltage probes with full spin mixing.  
The measured Hall voltage is directly related to spin current flowing 
in the system. 
An asymmetry between the upper and lower edges, e.g., 
introduced by removing voltage probe 3 or by gating, creates spin filtering effect: an unpolarized current injected
from source 1 induces a spin-polarized current flowing into drain 4.
Hall probes 5 and 6 downstream can serve as detectors of
spin currents.}
\label{fig4}
\end{figure}

Since the purpose of this section is mostly illustrative, we will keep 
our discussion as simple as possible. 
In particular, we shall ignore
transport in the bulk, leaving the discussion of its role for Sec.IV.
We also first neglect 
spin flip backscattering between edge states within one edge. 
Estimates of the spin flip rate wil be given below, 
Eq.(\ref{eq:SO_backscattering}).
A general approach, based on the Landauer-B\"uttiker formalism\,\cite{Buttiker},
which can be used to calculate spin and charge currents at the edge 
for any configuration of current and voltage leads, was presented 
in Ref.\cite{Abanin06a}.

In this approach, 
transport is described by a scattering matrix\,\cite{Buttiker}, 
with  the edge states
playing the role of scattering channels, and the reservoirs supplying
in-states and absorbing out-states. Current in each mode is described
by the relation $I_{\uparrow(\downarrow)}=\frac{e^2}{h}\phi_{\uparrow(\downarrow)}$, 
where $\phi_{\uparrow(\downarrow)}$ is the reservoir chemical potential
for given spin projection.
We consider the Hall bar geometry with four contacts 1-4 (see Fig.\ref{fig4}),
where the contacts 1 and 4 serve as current source and drain.
For these two contacts we do not assume spin mixing, 
so that the injected and drained current may be spin polarized.
The contacts 2, 3 are voltage probes, 
which means that they do not drain current from the system.
Furthermore, we assume that the probes provide full spin mixing,
i.e. 
chemical potentials of outgoing spin-up and spin-down electrons 
are equal. 

The simplest situation arises when unpolarized current is injected 
through contact 1. Then the up- and down-spins spatially separate 
in a symmetric way, flowing along the opposite edges of the bar.
This can be interpreted as circulating spin current,
and described as spin-Hall effect with quantized spin conductance
$\sigma_{xy}^{\rm (spin)}=e^2/h$. No electric voltage
will be induced between
the voltage probes 2, 3 in this case (zero charge-Hall effect).

This device can be used as a detector of spin polarized current,
made possible by the reciprocal of the spin Hall effect, 
in which the electric Hall voltage 
is directly related to spin rather than charge current. 
Suppose the up-spin and down-spin electrons, injected 
through contact 1, have unequal chemical potentials, $\phi_{\uparrow}\ne\phi_{\downarrow}$. 
Then the currents flowing into the probes 2 and 3, 
$I_{\uparrow(\downarrow)}=\frac{e^2}{h}\phi_{\uparrow(\downarrow)}$,
after equilibration and spin mixing in the probes,  
induce voltages 
$V_{2(3)}=\phi_{\uparrow(\downarrow)}/2$.
The resulting Hall voltage $V_{xy}=
(\phi_{\downarrow}-\phi_{\uparrow})/2$
is directly proportional to {\it spin current}.
At the same time, an unpolarized current (for which 
$\phi_{\downarrow}=\phi_{\uparrow}$) flows symmetrically in the 
upper and lower edges without generating Hall voltage.

Spin transport at $\nu=0$ also allows to realize spin filter. 
Suppose that the upper and lower edges of the device in Fig.\ref{fig4}
are made asymmetric,
which can be achieved, for example, simply by removing probe 3.
Then we inject unpolarized current $2I$ through contact 1. 
The injected current will be distributed equally between the upper and lower edges in cross section A. In cross section B, however,
the net current 
will be spin polarized due to spin
mixing in probe 2. The down-spin current reaching the drain
in the upper edge 
equals $I/2$ while 
the up-spin current in the lower edge is $I$. 
Therefore, the total drained current becomes spin 
polarized. 
The spin polarized current can be fed into 
another system (see Fig.\ref{fig4}), 
where it can be detected using Hall probes 5 and 6 as discussed above. 

More complicated circuits can be assembled which generate spin currents and 
detect them elsewhere. Note that the important principle is that 
as long as backscattering is not allowed, the edge current can travel long 
distances and the circuit is nonlocal, just as in the integer 
QHE\,\cite{Buttiker}. In this case the current-voltage 
relationship is obtained by solving the circuit equations as described in 
Ref.\cite{Abanin06a}. 
The spatial scale of nonlocality is controlled by spin relaxation which
can be due to spin-orbit interaction or due to magnetic impurities 
near graphene edge. 

For simplicity, here we limit the discussion to the effects of spin-orbit.
There are two main spin-orbit terms 
in the graphene Hamiltonian\,\cite{KaneMele,Min06}, the so-called intrinsic 
and Rashba interaction, given by
%
%
\be\label{eq:spin-orbit}
H_{\rm SO}=\lambda_{\rm SO}\sigma_z\tau_z s_z,\quad H_{\rm R}=\lambda_{\rm R}\left(\sigma_x \tau_z
 s_y-\sigma_y s_x \right), 
\ee
where Pauli matrices $\sigma_i$ act in the sublattice space (Dirac spinor), 
while $\tau_i$ act in the valley space, and $s_i$ represent physical spin. 
Estimates from band calculations\,\cite{Min06} give 
$\lambda_{\rm R}\approx 0.1 \,{\rm K}$ and a negligibly small 
$\lambda_{\rm SO}\approx 6 \,{\rm mK}$.

To estimate the backscattering rate 
due to the spin-orbit interaction (\ref{eq:spin-orbit}), we note that
for an ideal atomically sharp edge the spin-orbit would couple 
the left and right states with the same momentum, opening a minigap 
at branch crossing:
$\tilde\epsilon=\pm\sqrt{\epsilon^2+{\lambda_{\rm R}}^2}$.
However, this momentum-conserving interaction alone cannot 
backscatter edge states, 
and we need to take disorder
into account. Edges of graphite monolayers have been
imaged using STM probes\,\cite{Kobayashi06,Niimi06a},
where it was found that typically edge disorder can 
be viewed as patches of missing atoms of characteristic 
size $d\sim 1\,{\rm nm}$. Taking into account the left-right branch
mixing by spin-orbit $H_{\rm R}$, 
characterized by small mixing ratio of $\lambda_{\rm R}/|\epsilon|\ll1$ 
away from branch crossing, 
we obtain an estimate of the backscattering mean free path:
\be\label{eq:SO_backscattering}
\ell(\epsilon) \sim (\epsilon/\lambda_{\rm R})^2(\ell_B/d)^2 d
,\quad
|\epsilon|\gtrsim \lambda_{\rm R}
,
\ee
which gives $\ell\sim 10\,{\rm \mu m}$ for typical $\epsilon\sim 10\,{\rm K}$.
The factor $(\ell_B/d)^2\gtrsim1$ accounts for 
the magnetic field dependence of disorder matrix elements.


The quadratic energy dependence in (\ref{eq:SO_backscattering}),
with spin flip rate having a sharp peak near branch crossing,
suggests\,\cite{Abanin06a} the possibility to control backscattering 
using local gate. By tuning local chemical potential to and from the
branch crossing, where the spin flip rate has a sharp peak, 
Eq.(\ref{eq:SO_backscattering}), we can
induce or suppress backscattering in a controlled way. 
Spin filtering is achieved by controlling local
gates on opposite sides of the Hall bar asymmetrically.

\section{IV. Edge and bulk transport at $\nu=0$.}
\label{sec4}

Spin flip backscattering (\ref{eq:SO_backscattering}) can be incorporated
in the edge transport model, described by coupled equations 
for particle density in the two spin-polarized modes:
%
\be\label{eq:edge_12}
\begin{array}{l}
\partial_t n_1 + \partial_x \phi_1 = \gamma (\phi_2-\phi_1)
\\
\partial_t n_2 - \partial_x \phi_2 = \gamma (\phi_1-\phi_2) ,
\quad
n_i=\nu_i\phi_i
,
\end{array}
\ee
where 
$\gamma^{-1}=\ell$ is the backscattering mean free path (\ref{eq:SO_backscattering}) taken for $\epsilon$ at the Fermi level,
and $\nu_{1,2}$ are compressibilities
of the modes.  (For brevity, we use 1 and 2 instead of $\uparrow$ and 
$\downarrow$.) In writing Eqs.(\ref{eq:edge_12}) 
we implicitly assume
that fast energy relaxation maintains local equilibrium of each of the modes,
which is consistent with metallic temperature dependence of transport 
coefficients\,\cite{Abanin07}.

In a stationary state, Eqs.(\ref{eq:edge_12}) have an integral
$\tilde I=\phi_1-\phi_2$ which expresses current conservation at the edge.
[In this section we use the units of $e^2/h=1$.]
The general solution in the stationary current-carrying state is 
\be\label{eq:-Ex}
\phi_{1,2}(x)=\phi_{1,2}^{\ast}-\E x
,\quad
\E=\gamma \tilde I
\ee
Taking into account that $\tilde I$ is the current in one edge, we calculate 
the total current as 
\be\label{eq:edge_Rxx}
I=2\tilde I = \frac{2}{\gamma}\E
\ee
To describe the longitudinal resistance
in the four-terminsl geometry, one must add potential drop
on voltage probes\,\cite{Abanin07}, 
which gives $R=\frac12(\gamma L+1)$, where $L$ is the 
distance between the probes. Comparing to the data for $\rho_{xx}$ at $\nu=0$
we estimate\,\cite{Abanin07} $\ell\approx 0.5\,{\rm \mu m}$. 
This mean free path value, which is relatively small on the scale 
predicted by Eq.(\ref{eq:SO_backscattering}), 
can be explained if spin flip processes are dominated by nonintrinsic effects, 
such as magnetic impurities localized near the edge.

It is crucial that the edge transport model 
(\ref{eq:edge_12}) treats
both edges of a Hall bar in an identical way, 
thus predicting zero Hall effect.
In order to understand the observed density 
dependence of Hall coefficient\,\cite{Zhang06,Abanin07},
which changes sign smoothly at $\nu=0$ without exhibiting a plateau,
and of $\rho_{xx}$ which has a sharp peak at $\nu=0$, we need  
to incorporate transport in the bulk in our model.
In the full edge+bulk model, the density dependence of transport coefficients
arises from bulk currents short-circuiting edge currents away from
$\nu=0$. This explains, as we shall now see, the Hall effect, the
peak of $\rho_{xx}$, 
the resistance fluctuations near $\nu=0$, as well as the behavior
of $\sigma_{xx}$ and $\sigma_{xy}$.

We describe the transport problem in the bulk by the current-field relation, 
separately for each spin projection:
\[
\vec j_i =-\hat\sigma_i\nabla\psi_i
,\quad
\hat\sigma_i=\lp\matrix{\sigma_{xx}^{(i)} &\sigma_{xy}^{(i)}\cr -\sigma_{xy}^{(i)} &\sigma_{xx}^{(i)}}\rp
,\quad
i=1,2,
\]
where $\psi_{1,2}$ are electrochemical potentials for two spin states.
We assume that 
the bulk conductivities $\sigma_{xx}^{(1,2)}$, as a function
of density $\nu$, are peaked at the spin-split Landau levels. 
For simplicity, here we ignore possible valley splitting, 
in which case the spin up and down Landau levels occur at $\nu=\pm1$
around the Dirac point. 
As a simplest model, below we use Gaussians
\be\label{eq:gaussians}
\sigma^{(1)}_{xx}(\nu)=e^{-A(\nu-1)^2}
,\quad
\sigma^{(2)}_{xx}(\nu)=e^{-A(\nu+1)^2}
\ee
with  the parameter $A$ describing the width of the levels. 
The Hall conductivities $\sigma_{xy}^{(1,2)}$ exhibit plateaus
on either side of the peak in $\sigma_{xx}^{(1,2)}$. The dependence
of $\sigma_{xy}^{(1,2)}$ on $\nu$ can be modeled with the help of the 
semicircle relation
$\sigma^{(1,2)}_{xy}(\sigma^{(1,2)}_{xy}\mp 2)+(\sigma^{(1,2)}_{xx})^2=0$
which often provides a good 
description of conventional QHE systems\,\cite{semicircle_relation}.

The condition of charge continuity, $\nabla\vec j_i =0$, gives a 2d Laplace's
equation for the potentials, $\nabla^2\psi_i=0$. 
This equation must be solved together 
with the boundary conditions phenomenologically describing bulk-edge coupling:
\be\label{eq:bound_cond}
\vec n.\vec j_i = g(\psi_i-\phi_i)
\ee
where $\vec n$ is a normal vector to the boundary,
and $g(\psi_i-\phi_i)$
represents the edge-bulk leakage current density.

Although a general solution of this problem can be given with the help of Fourier method, here we consider only the case when the potentials 
$\psi_i(\vec x)$ are varying slowly 
on the scale of the bar width $w$, 
which will suffice for our analysis of a homogeneous 
current flow. In this case, linearizing $\psi_i(\vec x)$ in the direction
transverse to the bar, we can write 
Eqs.(\ref{eq:bound_cond}) for both edges of the bar as
\be\label{eq:bulk}
\begin{array}{l}
 -\sigma_{xy}\p_x \psi_i+\sigma_{xx}(\psi_{i'}-\psi_i)/w = g(\psi_i-\phi_i)
\\
\sigma_{xy}\p_x \psi_{i'}+\sigma_{xx}(\psi_i-\psi_{i'})/w = g(\psi_{i'}-\phi_{i'})
,
\end{array}
\ee
$i=1,2$, where the primed and unprimed quantities 
denote variables at opposite edges of the bar.

Equations for the edge variables $\phi_i$ are obtained by 
adding the bulk-edge leakage term to Eqs.(\ref{eq:edge_12}), giving
\be\label{eq:edge_12_g}
\begin{array}{r}
\p_x \phi_1 = \gamma(\phi_2-\phi_1)+g(\psi_1-\phi_1),
\\
-\p_x \phi_2 = \gamma(\phi_1-\phi_2)+g(\psi_2-\phi_2),
\end{array}
\ee
along with a similar pair of equations for $\phi_{1'}$, $\phi_{2'}$ at the opposite edge.

The solution of these eight equations, describing uniform current,
is of the form 
$\phi_i=\phi_i^{\ast}-\E x$, $\psi_i=\psi_i^{\ast}-\E x$, etc.,
with the same linear part $-\E x$ for all quantities. Using the algebraic structure
of this linear system and the symmetry between the 
edges, we reduce the number of equations from eight to two. 
First, it is convenient to
express the parameters 
$\phi_i^{\ast}$ through $\psi_i^{\ast}$ using Eqs.(\ref{eq:edge_12_g}),
which gives
\be\label{eq:phi-psi}
\begin{array}{r}
\phi_1^{\ast}=\frac{\gamma+g}{2\gamma+g}\psi_1^{\ast}+\frac{\gamma}{2\gamma+g}\psi_2^{\ast}
+\frac{\E}{2\gamma+g}
\\
\phi_2^{\ast}=\frac{\gamma+g}{2\gamma+g}\psi_2^{\ast}+\frac{\gamma}{2\gamma+g}\psi_1^{\ast}
-\frac{\E}{2\gamma+g}
\end{array}
\ee
Writing similar equations for the variables at the opposite edge to express
$\phi_{1'}$, $\phi_{2'}$ through $\psi_{1'}$, $\psi_{2'}$, and substituting the result 
in Eqs.(\ref{eq:bulk}), we obtain four equations for 
$\psi_i$ and $\psi_{i'}$ which have the form
\be\label{eq:four_eqs}
\begin{array}{r}
 -\tilde\sigma_{xy}^{(1)}w\E =\sigma_{xx}^{(1)}(\psi_{1'}^{\ast}-\psi_1^{\ast})+\lambda(\psi_2^{\ast}-\psi_1^{\ast})
\\
 \tilde\sigma_{xy}^{(1)}w\E =\sigma_{xx}^{(1)}(\psi_1^{\ast}-\psi_{1'}^{\ast})+\lambda(\psi_{2'}^{\ast}-\psi_{1'}^{\ast})
\\
 -\tilde\sigma_{xy}^{(2)}w\E =\sigma_{xx}^{(2)}(\psi_{2'}^{\ast}-\psi_2^{\ast})+\lambda(\psi_1^{\ast}-\psi_2^{\ast})
\\
 \tilde\sigma_{xy}^{(2)}w\E =\sigma_{xx}^{(2)}(\psi_2^{\ast}-\psi_{2'}^{\ast})+\lambda(\psi_{1'}^{\ast}-\psi_{2'}^{\ast})
\end{array}
\ee
where the coefficients in this linear system are defined as
\be
\tilde\sigma_{xy}^{(1,2)}=\sigma_{xy}^{(1,2)}\pm\frac{g}{2\gamma+g}
,\quad
\lambda=\frac{w\gamma g}{2\gamma+g}
.
\ee
The quantities 
$\tilde\sigma^{(1,2)}_{xy}$ represent the sum 
of the bulk and edge contributions to Hall conductivity for each spin.

Symmetry between the edges allows to further reduce 
the number of independent variables.
For that we add the first two equations to obtain
$\psi_1^{\ast}+\psi_{1'}^{\ast}=\psi_2^{\ast}+\psi_{2'}^{\ast}$. Also we note that
all potentials can be changed by the same constant that can be chosen so that
the new quantities $\psi_i^{\ast}$ and $\psi_{i'}^{\ast}$ satisfy
$\psi_i^{\ast}=-\psi_{i'}^{\ast}$.
After that Eqs.(\ref{eq:four_eqs}) yield
\be\label{eq:two_eqs}
\begin{array}{r}
\tilde\sigma_{xy}^{(1)}w\E =2\sigma_{xx}^{(1)}\psi_1^{\ast}-\lambda(\psi_2^{\ast}-\psi_1^{\ast})
\\
\tilde\sigma_{xy}^{(2)}w\E =2\sigma_{xx}^{(2)}\psi_2^{\ast}-\lambda(\psi_1^{\ast}-\psi_2^{\ast})
\end{array}
\ee
These two equations can be solved to find $\psi_{1,2}^{\ast}$.

Now we can find the current as a sum of the edge and bulk contributions,
$I=I_{\rm edge}+I_{\rm bulk}$, where
\[
I_{\rm edge}=\phi_1-\phi_2+\phi_{2'}-\phi_{1'}=2(\phi_1^{\ast}-\phi_2^{\ast})
\]
%
and
\[
I_{\rm bulk}=\sigma_{xy}^{(1)}(\psi_1-\psi_{1'})+\sigma_{xx}^{(1)}w\E
+\sigma_{xy}^{(2)}(\psi_2-\psi_{2'})+\sigma_{xx}^{(2)}w\E
\]
After expressing $\phi_i$ through $\psi_i$ with the help 
of Eqs.(\ref{eq:phi-psi}) and using the solution of Eqs.(\ref{eq:two_eqs}), 
we obtain a relation $I=2{\cal E}/\tilde\gamma$, where
\be\label{eq:G_total}
\frac2{\tilde\gamma}
=\frac4{2\gamma+g}+ \frac{w}{\rho^{(1)}_{xx}}+\frac{w}{\rho^{(2)}_{xx}}
-\frac{\lambda w\lp \tilde\sigma^{(1)}_{xy}/\sigma^{(1)}_{xx}-\tilde\sigma^{(2)}_{xy}/\sigma^{(2)}_{xx}\rp^2}{2+\lambda/\sigma^{(1)}_{xx}+\lambda/\sigma^{(2)}_{xx}}.
\ee
The quantities $\rho^{(1,2)}_{xx}$ are defined as
$\rho^{(i)}_{xx}=\sigma^{(i)}_{xx}/({{\tilde\sigma}^{(i)}_{xy}}{}^2+\sigma^{(i)}_{xx}{}^2)$.
The quantity $\tilde\gamma$, Eq.(\ref{eq:G_total}), replaces $\gamma$ in Eq.(\ref{eq:edge_Rxx}).
In the absence of bulk conductivity, $\sigma^{(1,2)}_{xx}\to 0$,
we recover the result for pure edge transport, $\tilde\gamma=\gamma$.

The Hall voltage can be calculated from this solution as 
$V_{H}=\frac12(\phi_1+\phi_2-\phi_{1'}-\phi_{2'})$, where $\phi_i$, $\phi_{i'}$
are variables at opposite edges. We obtain $V_{H}=\xi{\cal E}$, where
\be\label{eq:G_H}
\xi=2w\frac{\tilde\sigma^{(1)}_{xy}\lp \lambda+\sigma^{(2)}_{xx}\rp
+\tilde\sigma^{(2)}_{xy}\lp \lambda+\sigma^{(1)}_{xx}\rp}{2\sigma^{(1)}_{xx}\sigma^{(2)}_{xx}+\lambda\sigma^{(2)}_{xx}+\lambda\sigma^{(1)}_{xx}}
.
\ee
This quantity vanishes at $\nu=0$, since 
$\sigma^{(1)}_{xy}=-\sigma^{(2)}_{xy}$ and 
$\sigma^{(1)}_{xx}=\sigma^{(2)}_{xx}$ at this point
due to particle-hole symmetry.

\begin{figure}
\includegraphics[width=3.3in]{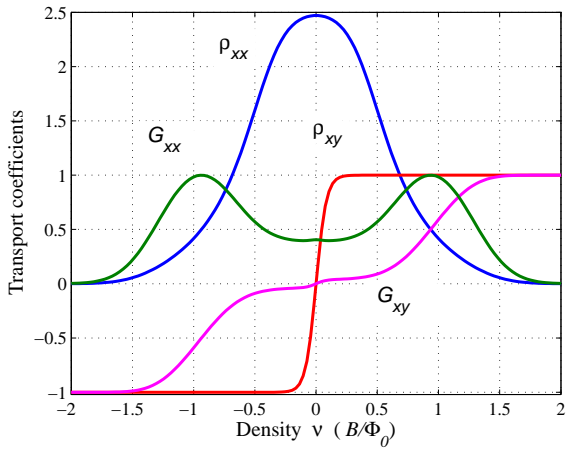}
\caption[]{
Density dependence of transport coefficients $\rho_{xx}=\tilde\gamma w/2$, $\rho_{xy=\tilde\gamma \xi/2}$
and $G_{xx}=\rho_{xy}/(\rho_{xy}{}^2+\rho_{xx}{}^2)$, 
$G_{xy}=\rho_{xy}/(\rho_{xy}{}^2+\rho_{xx}{}^2)$,
obtained 
from the 
edge transport model (\ref{eq:edge_12_g}) augmented with
bulk conductivity, Eqs.(\ref{eq:bulk})
(see Eqs.(\ref{eq:G_total}),(\ref{eq:G_H}) and text). 
Parameter values: $A=6$, $\gamma w=5$.
Note the peak in $\rho_{xx}$,
the smooth behavior of $\rho_{xy}$ near $\nu=0$,
a quasi-plateau in  
$G_{xy}$, and a double-peak structure in $G_{xx}$.
}
\label{fig_transport}
\end{figure}

Transport coefficients, obtained from this model for typical parameter values, are displayed in
Fig.\ref{fig_transport} which reproduces many of the key features of the data 
(see Fig.1 in Ref.\cite{Abanin07}). 
In particular, the peak in $\rho_{xx}$ is due to edge transport
near $\nu=0$. The suppression of $\rho_{xx}$ at finite $\nu$ is due to
the bulk conductivity short-circuiting the edge transport.
The bulk and edge contributions to transport can be discerned from the
double peak structure in $G_{xx}$ in
Fig.\ref{fig_transport}. The peaks correspond to the bulk Landau level
contributions, Eq.(\ref{eq:gaussians}), whereas the part of 
$G_{xx}$ between the peaks, exceeding the superposition of two Gaussians,
Eq.(\ref{eq:gaussians}), is the edge contribution.
   The Hall resistance $\rho_{xy}$ is nonzero due to imbalance in
$\sigma_{xy}^{(1,2)}$ for opposite spin polarizations away from $\nu=0$. 
Interestingly, $\rho_{xy}$ in Fig.\ref{fig_transport} exhibits no plateau, 
while $G_{xy}$ calculated from $\rho_{xy}$ and $\rho_{xx}$ 
displays an under-developed plateau-like feature. 
Overall, this behavior resembles that of the experimentally measured
transport coefficients\,\cite{Zhang06,Abanin07}.

Another notable feature of the measured $\rho_{xy}$ and $\rho_{xx}$
is enhanced fluctuations near zero $\nu$. These fluctuations are
found to be strong in  Ref.\cite{Zhang06}, 
where $\rho_{xy}$ changes sign several times
near $\nu=0$. They are also present, although are not
as dramatic, in Ref.\cite{Abanin07}.
In the latter case, both $\rho_{xy}$ and $\rho_{xx}$ exhibit noisy behavior
in the interval near $\nu=0$ comparable to the $\rho_{xx}$ peak width.
As Ref.\cite{Abanin07} points out, this behavior is consistent with the 
edge transport model. In the absence of bulk transport,
the distribution of potential along the edge depends on 
the local backscattering rate $\gamma(x)$, whereby
Eq.(\ref{eq:-Ex}) is replaced by
\[
\phi_{1,2}(x)=\phi_{1,2}(0)-\tilde I\int_0^x\gamma(x')dx'
.
\]
Fluctuations of $\gamma$ arise due to its sensitivity to
the local value of Fermi energy in the spin-orbit scattering model,
Eq.(\ref{eq:SO_backscattering}), and, similarly,
for the magnetic impurity scattering mechanism.
Assuming that the random part of $\gamma$ is of a white noise character,
we obtain strong fluctuations $\delta\phi_{1,2}(x)$ 
along the edge of magnitude that
scales as a square root of the edge length.
These fluctuations will contribute equally 
to the longitudinal and transverse 
voltage, since they are uncorrelated on the
opposite sides of the Hall bar. The absence of fluctuations away from $\nu=0$
can be understood as a result of bulk conductivity short-circuiting the
edge current, which will equilibrate potentials on the opposite sides of the 
Hall bar.

The above discussion summarizes the results drawn from an attempt to model
quantum Hall transport in graphene at $\nu=0$ 
by counter-circulating edge states. 
By taking into account
backscattering within one edge as well as conduction
in the bulk which 
short-circuits edge transport away from the neutrality point,
this model accounts for the observed behavior of transport coefficients.
Still, since no direct evidence for spin polarization has yet been
found, more experimental and theoretical work will be needed to confirm the
chiral spin-polarized edge picture of the $\nu=0$ state. 
If proven to exist in graphene, these states will provide a unique setting to study spin transport as well as other interesting phenomena.

This work is supported by NSF MRSEC Program (DMR 02132802),
NSF-NIRT DMR-0304019 (DA, LL), and NSF grant DMR-0517222 (PAL).


\end{document}